\documentclass[12pt]{article}
\usepackage[utf8]{inputenc}
\usepackage[english]{babel}
\usepackage{epsfig,amsfonts,amsthm}
\usepackage{amsmath,amssymb}
\usepackage{xcolor}
\usepackage{cite}

\newcommand{\be}{\begin{equation}}
\newcommand{\ee}{\end{equation}}
\newcommand{\bea}{\begin{eqnarray}}
\newcommand{\eea}{\end{eqnarray}}

\renewcommand{\Re}{\mathrm{Re}\,}

\newcommand{\lr}[1]{ \langle #1 \rangle}

\newcommand{\mmmatrix}[9]{ \left(\! \begin{array}{ccc}#1 & #2 & #3\\ #4 & #5 & #6\\ #7 & #8 & #9\\ \end{array}\!\right) }
\newcommand{\toCP}{\xrightarrow{CP}}

\newcommand{\id}{\mathbf{1}}

\def\lsim{\mathrel{\rlap{\lower4pt\hbox{\hskip1pt$\sim$}}
    \raise1pt\hbox{$<$}}}         
\def\gsim{\mathrel{\rlap{\lower4pt\hbox{\hskip1pt$\sim$}}
    \raise1pt\hbox{$>$}}}         

\title{Constraining CP4 3HDM with top quark decays}

\author{Igor~P.~Ivanov$^{1}$\thanks{E-mail: ivanov@mail.sysu.edu.cn}, Semyon A. Obodenko$^{2}$\thanks{E-mail: semeon.obodenko@gmail.com}   
	\\
	{\small $^1$ School of Physics and Astronomy, Sun Yat-sen University, 519082 Zhuhai, China}\\
	{\small $^2$ Institute of Physics, Kazan Federal University, 16a Kremlyovskaya St., 42000 Kazan, Russia}
}

\begin{document}
	
	\maketitle
	
	\bigskip
	\begin{abstract}
CP4 3HDM is a unique three-Higgs-doublet model equipped with a higher-order $CP$ symmetry in the scalar and Yukawa sector.
Based on a single assumption (the minimal model with a $CP$-symmetry of order 4 and no accidental symmetry),
it leads to a remarkable correlation between its scalar and Yukawa sectors, which echoes in its phenomenology.
A recent scan of the parameter space of CP4 3HDM under the assumption of scalar alignment
identified a few dozens of points which passed many flavour constraints.
In the present work we show, however, that almost all of these points are now ruled out by the recent
LHC searches of $t \to H^+ b$ with subsequent hadronic decays of $H^+$.
Apart from a few points with charged Higgses heavier than the top quark,
only one point survives all the checks, the model with an exotic, non-2HDM-like generation pattern of $H^+$ couplings with quarks.
One can expect many more points with exotic $H^+$ couplings to quarks if the scalar alignment assumption is relaxed.
	\end{abstract}

\section{Introduction}

The Standard Model (SM), despite its lasting success,
does not provide any clue on the origin of the fermion masses, their mixing patterns,
and $CP$ violation.
Many models beyond the SM (bSM) address these issues by postulating new fields, symmetries, or interactions.
A conservative approach is to assume that, around the electroweak energy scale,
bSM physics manifests itself only via a non-minimal Higgs sector.
Although the LHC measurements show that the discovered 125 GeV Higgs is very SM-like 
\cite{Aad:2012gk,Chatrchyan:2012gu,Khachatryan:2016vau},
it is well possible that there exist (many) other scalar fields, 
which have escaped collider searches due to decoupling from the main Higgs production channels.

Among various options for non-minimal Higgs sectors \cite{Ivanov:2017dad,Kanemura:2014bqa},
$N$-Higgs-doublet models (NHDM) remain among the simplest and most attractive frameworks.
Based on the simple idea that Higgs doublets can come in generations,
NHDM can accommodate various new symmetries, discrete \cite{Ivanov:2012ry,Ivanov:2012fp} or continuous \cite{Darvishi:2019dbh},
with phenomenologically interesting consequences.

The two-Higgs-doublet model (2HDM), by far the most explored version, has had limited success 
in linking the observed fermion properties with patterns of the scalar sector \cite{Branco:2011iw}. 
Very few symmetries are possible with two Higgs doublets
\cite{Ivanov:2006yq,Nishi:2006tg,Ivanov:2007de,Maniatis:2007de,Ferreira:2009wh,Ferreira:2010ir}.
Although they can be used to shape the Yukawa sectors 
(see the recent developments \cite{Cogollo:2016dsd,Alves:2018kjr,Nebot:2020niz} beyond the classical 2HDM Type-I and Type-II models), 
they are insufficiently constraining or they predict highly peculiar properties not seen in experiment 
\cite{Maniatis:2009vp,Ferreira:2010ir}.

The three-Higgs doublet model (3HDM) can accommodate many more symmetry groups \cite{Ivanov:2012ry,Ivanov:2012fp,Darvishi:2019dbh},
some of which were used back in early 1980's to deduce the fermion masses and mixing hierarchies from symmetries,
see a historical overview in \cite{Ivanov:2017dad}.
Despite significant efforts, the overall situation is similar to the 2HDM. 
If one imposes a large discrete symmetry group, which would strongly constrain the scalar and Yukawa sectors, 
one predicts features which are in conflict with experiment, such as massless or mass-degenerate quarks,
insufficient mixing, or the absence of $CP$-violation.
The pedagogical insight into the anatomy of this mismatch in $A_4$ and $S_4$-symmetric 3HDMs
was given in \cite{Felipe:2013ie}. In \cite{Felipe:2014zka}, improving the earlier observations of \cite{Leurer:1992wg},
it was shown that this failure stems from inability of sufficiently large symmetry groups
to break down completely upon minimization of the Higgs potential.
If, instead, one builds a 3HDM with a small symmetry group, or if one assumes that a symmetry of softly broken,
one can successfully reproduce fermion masses and mixing at the expense of losing the symmetry control over the flavor properties.
Each sector (scalar, up quark, and down quark) comes with its own free parameters, and one typically loses predictivity.

\subsection{CP4 3HDM}

Recently, a version of 3HDM was proposed in \cite{Ivanov:2015mwl} which combines the minimality of assumptions with 
a surprising degree of control over the Yukawa sector.
The model is based on a single symmetry, the $CP$-symmetry of order 4 (CP4), without any other accidental symmetry.
Despite not respecting the usual $CP$ symmetry, the model is nevertheless $CP$-conserving 
\cite{Ivanov:2015mwl,Aranda:2016qmp,Haber:2018iwr}.
It also offers the first example of a {\em physically distinct} form of $CP$-symmetry, with observables
specific only to CP4 \cite{Haber:2018iwr}, 
and therefore it can in principle be distinguished from all other $CP$-conserving models constructed so far.

The CP4 symmetry was extended to the quark Yukawa sector in \cite{Ferreira:2017tvy},
leading to unusually strong constraints on the Yukawa matrices.
In order to avoid mass-degenerate quarks, the initial CP4 symmetry must be spontaneously broken,
which can be easily arranged by the scalar potential.
After symmetry breaking, the Yukawa sector still contains enough free parameters to accommodate
quark masses and mixing, as well as the appropriate amount of $CP$ violation.
The numerical scan performed in \cite{Ferreira:2017tvy} gave many points in the parameter space 
which satisfy the electroweak precision tests and do not violate 
the kaon and $B$-meson oscillation parameters.
After that, there remains very little freedom, and the model exhibits correlations which can be tested 
with further experimental results.

In short, the parameter space scans performed in \cite{Ferreira:2017tvy} demonstrated that 
the CP4 3HDM can accomplish unexpectedly much for a 3HDM based on a single symmetry.
Therefore, it is interesting to check if additional phenomenological constraints can be satisfied in this minimalistic setting.

\subsection{Light charged Higgses in top quark decays}

Due to the intrinsic relations between the scalar and Yukawa sectors of the CP4 3HDM, 
one expects scalar-induced flavor-changing neutral currents (FCNC).
The FCNC generated by the SM-like Higgs can be eliminated by the simplifying assumption of an exact alignment in the scalar sector,
the approach used in \cite{Ferreira:2017tvy}. However one still expects
FCNCs in additional neutral Higgses, as well as non-trivial generation patterns in the charged Higgs couplings with quarks. 

In this paper, we investigate a particular feature which emerged from the numerical scan of \cite{Ferreira:2017tvy}.
The majority of the parameter space points which passed the constraints of \cite{Ferreira:2017tvy} 
led to one or both charged Higgses $H_1^+$ and $H_2^+$ being lighter than the top quark.
This allows for the decays $t\to H_{1,2}^+ d_i$, where $d_i = (d, s, b)$, with the subsequent
decays of the charged Higgses $H_{1,2}^+ \to u_i \bar{d}_j$, with $u_i = (u, c, s)$.
Such decays have been studied theoretically in the 2HDM \cite{Branco:2011iw,Akeroyd:2016ymd,Arbey:2017gmh}
as well as in various versions of the 3HDM equipped with natural flavor conservation (NFC) 
\cite{Grossman:1994jb,Akeroyd:2012yg,Akeroyd:2016ssd,Akeroyd:2018axd,Akeroyd:2019mvt,Chakraborti:2021bpy}.
These decays were also searched for at the LHC \cite{Aad:2013hla, Khachatryan:2015uua}. 
After Ref.~\cite{Ferreira:2017tvy} appeared, two more searches were published 
by the CMS collaboration \cite{Sirunyan:2018dvm,Sirunyan:2020aln}.
All of them gave negative results, with the upper limits on the corresponding branching ratios at the subpercent level
(see details below).

These searches required an isolated lepton, a large missing $E_T$, 
and the presence of at least two $b$-jets indicative of the $t\bar t$ production.
To separate the signal from the SM background, at least one additional heavy-quark jet, $b$ or $c$, 
arising from the $H^+ \to c\bar s$ or $H^+ \to c\bar b$ decays was required.
These $H^+$ decay channels naturally arise in 2HDM or 3HDMs with natural flavor conservation.
However, in CP4 3HDM, the flavor preferences of the off-diagonal couplings $\bar u_i d_j H_{1,2}^+$ 
do not necessarily follow the this pattern.
In particular, the largest off-diagonal couplings do not always link the heaviest fermions.
Thus, it remains an open question whether the existing ATLAS and CMS constraints rule out the CP4 3HDM examples
with light charged Higgses.
This is the question we address in the present paper.

Before going into the details, let us mention that charged Higgs bosons
can of course manifest themselves through other processes, such as the direct pair production, Drell-Yan process,
charged Higgs loop contributions to the $h_{SM}\to \gamma\gamma$ decay, to name a few. 
These signatures have been studied at length for the 2HDM \cite{Branco:2011iw,Akeroyd:2016ymd,Arbey:2017gmh}, 
and the negative results of their searches were used to constrain the 2HDM parameter space.
However, care should be taken when extending these constraints beyond the two-Higgs-doublet models,
in particular, to 3HDM. Just to give an example, it is well known that $B_s \to X\gamma$ decays
rule out charged Higgs bosons of the 2HDM Type II for masses below about 600 GeV \cite{Misiak:2017bgg,Arbey:2017gmh}.
However, this limit is much weaker for the 2HDM Type I \cite{Misiak:2017bgg,Arhrib:2020tqk}, 2HDM Type III \cite{HernandezSanchez:2012eg},
for the so-called Aligned 2HDM \cite{Akeroyd:2012yg}, as well as
in models with more than two doublets equipped with NFC
\cite{Akeroyd:2012yg,Akeroyd:2016ssd,Chakraborti:2021bpy}.

Since CP4 3HDM does {\em not} possess the NFC property, it will be interesting to see, eventually, 
whether the light charged Higgses could be ruled out by $B_s \to X\gamma$ or other collider searches. 
This requires substantial work. What we point out in this paper is that a much more direct check can first be done
before plunging into the radiative meson decay machinery: comparing the CP4 3HDM predictions 
for the top quark decay chains involving charged Higgses with the LHC searches.
As we will show, a combination of three experimental checks rules out almost all the viable points
identified in \cite{Ferreira:2017tvy}. However, a detailed analysis also reveals a few highly exotic patterns
of charged Higgs interaction with quarks, which will be interesting to check in new scans of the CP4 3HDM parameter space.

The structure of the paper is the following. In the next section, we give a brief reminder
of the model CP4 3HDM and discuss the role of light charged Higgs bosons in top quark decays.
In Section~\ref{section-numerical} we study three observables: the total top quark decay width $\Gamma_t$
and the two decay chains investigated experimentally at the LHC:
$t \to H^+ b$ followed by $Br(H^+ \to c\bar s)$ or $Br(H^+ \to c\bar b)$.
We show that almost all points of \cite{Ferreira:2017tvy} fail in at least one of these tests.
We discuss and summarize our results in the last section. An appendix provides some details
on the possible Yukawa sectors of the CP4 3HDM.
 

\section{Charged Higgses in CP4 3HDM} 

\subsection{CP4 3HDM scalar sector}

The 3HDMs make use of three Higgs doublets $\phi_i$, $i = 1,2,3$ with identical quantum numbers.
CP4 is a transformation which maps Higgs doublets to their conjugates with a simultaneous
rotation in the doublet space. Following \cite{Ivanov:2015mwl,Ferreira:2017tvy}, we use the following form of the CP4:
\begin{equation}
\phi_a \toCP X_{ab} \phi_b^*\,,\quad
X =  \left(\begin{array}{ccc}
	1 & 0 & 0 \\
	0 & 0 & i  \\
	0 & -i & 0
\end{array}\right)\,.
\label{CP4-def}
\end{equation}
Applying this transformation twices leads to the Higgs family transformation with the matrix 
$XX^* = {\rm diag}(1,-1,-1) \not = \id$. In order to get the identity transformation,
one must apply CP4 four times, hence order-4 transformation. 
It is known that any CP-type transformation of order 4 acting in the space of three complex fields
can be turned into \eqref{CP4-def} by a suitable basis change \cite{weinberg-vol1}.

The most general renormalizable 3HDM potential respecting this symmetry \cite{Ivanov:2015mwl} can be written as 
$V = V_0+V_1$ where
\begin{eqnarray}
V_0 &=& - m_{11}^2 (\phi_1^\dagger \phi_1) - m_{22}^2 (\phi_2^\dagger \phi_2 + \phi_3^\dagger \phi_3)
+ \lambda_1 (\phi_1^\dagger \phi_1)^2 + \lambda_2 \left[(\phi_2^\dagger \phi_2)^2 + (\phi_3^\dagger \phi_3)^2\right]
\nonumber\\
&+& \lambda_3 (\phi_1^\dagger \phi_1) (\phi_2^\dagger \phi_2 + \phi_3^\dagger \phi_3)
+ \lambda'_3 (\phi_2^\dagger \phi_2) (\phi_3^\dagger \phi_3)\nonumber\\
&+& \lambda_4 \left[(\phi_1^\dagger \phi_2)(\phi_2^\dagger \phi_1) + (\phi_1^\dagger \phi_3)(\phi_3^\dagger \phi_1)\right]
+ \lambda'_4 (\phi_2^\dagger \phi_3)(\phi_3^\dagger \phi_2)\,,
\label{V0}
\end{eqnarray}
with all parameters being real, and
\begin{equation}
V_1 = \lambda_5 (\phi_3^\dagger\phi_1)(\phi_2^\dagger\phi_1) +
\lambda_8(\phi_2^\dagger \phi_3)^2 + \lambda_9(\phi_2^\dagger\phi_3)(\phi_2^\dagger\phi_2-\phi_3^\dagger\phi_3) + h.c.
\label{V1}
\end{equation}
with real $\lambda_5$ and complex $\lambda_8$, $\lambda_9$. 

Minimization of this potential and the resulting scalar bosons mass matrices were studied in \cite{Ferreira:2017tvy}.
The minimum breaks CP4, thus leading to a spontaneously broken $CP$ symmetry.
Expansion of potential near the minimum produces five neutral scalar bosons and two pairs of charged Higgses $H^{\pm}_{1,2}$.
For a generic setting, all neutral Higgs bosons can couple to $WW$ and $ZZ$ pairs.
However, if one fixes $m_{11}^2 = m_{22}^2$, the model displays the scalar alignment property: one of the neutral Higgses
$h$ couples to the $WW$ and $ZZ$ exactly as in the SM, while the other four neutral boson decouple from these channels.
In this case, the additional Higgses cannot be produced in gauge-boson fusion, and the only way 
to produce them would be through their couplings to quarks,
which may help these bosons escape the present day LHC searches.

A particular feature which emerged from the numerical scan of \cite{Ferreira:2017tvy} was that the additional Higgses turned out rather light,
with the masses of few hundred GeV. Thus, this particular version of the model {\em does not} possess the decoupling limit \cite{Gunion:2002zf}
(in fact, the results of \cite{Carrolo:2021euy} indicate that a spontaneously broken CP4 3HDM cannot possess this limit).
Even more impressive, almost all parameter space points emerging from the numerical scan of \cite{Ferreira:2017tvy}
contained one or two charged Higgses lighter than the top quark.
This feature opens up new channels for the top decay to light quarks and the charged Higgses $H^\pm_{1,2}$, 
with subsequent quark decays of $H_{1,2}^\pm$. 

These channels offers a robust check of the model.
Indeed, we do not need to estimate direct production of the charged Higgses in $q\bar q$ collisions at the LHC,
which may be subject to uncertainties. We simply need to analyze the top decay properties
and search for $t\to d_j H_{1,2}^+ (\to u_i \bar{d}_k)$, where $d_j$ stands for $(d, s, b)$ and $u_i = (u, c)$.
However this signal may differ substantially from the 2HDM pattern because there is no guarantee that 
the preferred decay chains involve the heaviest quarks. Thus, it is not immediately clear 
whether light charged Higgses are in conflict with the LHC searches.

\subsection{CP4 3HDM Yukawa sector}

In order to describe charge Higgs coupling preferences, we briefly recapitulate 
the structure of the CP4 3HDM Yukawa sector, explored in detail in \cite{Ferreira:2017tvy}.
The quark Yukawa Lagrangian
\begin{equation}
-{\cal L}_Y = \bar q_L \Gamma_a d_R \phi_a + \bar q_L \Delta_a u_R \tilde\phi_a + h.c.,\label{Yukawa-general}
\end{equation}
where $\tilde\phi_a = i\sigma_2 \phi_a^* = (\phi_{a}^{0*}, - \phi_a^-)^T$
can be made CP4 invariant if we assume that CP4 acts non-trivially not only on the scalar doublets but also on fermions
\begin{equation}
\psi_i \toCP Y_{ij} \psi_j^{CP}\,, \quad\mbox{where} \quad \psi^{CP} = \gamma^0 C \bar\psi^T\,.
\label{fermion-GCP}
\end{equation}
The Yukawa matrices $\Gamma_a$ and $\Delta_a$
can only be of special types, producing cases $A$, $B_1$, $B_2$, $B_3$ in the up and down quark sectors, see details in the Appendix.
These cases can be combined: one can pick up one case for the down quarks and another for the up quarks,
provided the left-handed doublets transform in the same way.
However, several combinations were ruled out since they induced a way too strong meson oscillations.
As a result, only three pairs were found possible in \cite{Ferreira:2017tvy}: cases $(B_1, B_1)$, $(B_2, B_2)$, and $(B_1, B_3)$,
for the down and up quarks, respectively.

Once the scalar potential and the Yukawa sector are constructed, one minimizes of the scalar potential and
obtains the vevs: $\lr{\phi_a^0} = v_a/\sqrt{2}$, which are in general complex.
One then substitutes them into the Yukawa sector and obtains the quark mass matrices:
\begin{equation}
\bar d_L M_d d_R + \bar u_L M_u u_L + h.c.,\quad
M_d = \frac{1}{\sqrt{2}}\sum_a \Gamma_a v_a\,, \quad M_u = \frac{1}{\sqrt{2}}\sum_a \Delta_a v_a^*\,.
\end{equation}
As usual, we switch to the physical quark fields, $d_{L} = V_{dL} d^{\rm phys.}_L$, $d_{R} = V_{dR} d^{\rm phys.}_R$,
$u_{L} = V_{uL} u^{\rm phys.}_L$, $u_{R} = V_{uR} u^{\rm phys.}_R$, so that the mass matrices $M_d$ and $M_u$ become diagonal.
The $\bar u_L d_L W^+$ interaction then becomes non-diagonal leading to the CKM matrix $V = V^\dagger_{uL} V_{dL}$.
The fitting procedure used in \cite{Ferreira:2017tvy} made sure that all quark masses and mixing parameters as well as 
the amount of $CP$ violation coincide with the experimental results.

\subsection{Charged Higgs bosons couplings}

We are interested here in finding the physical charged Higgs interactions with physical quarks. 
We use \eqref{Yukawa-general} to extract the charged scalar interaction matrices
\begin{equation}
-{\cal L}_{ch.} = (\bar u_L \Gamma_a d_R - \bar u_R \Delta^\dagger_a d_L) \phi_a^+ + h.c.\label{Yukawa-charged}
\end{equation}
Then we perform the rotations in the quark spaces as outlined above as well as the rotation in the charged scalars space:
\begin{equation}
\phi_a^+ = R_{ab} H_b^+\,,
\label{charged-Higgs-basis}
\end{equation}
where the index $b= 0,1,2$, so that $H_b^+ = (G^+, H_1^+, H_2^+)^T$,
with $b=0$ corresponding to the charged would-be Goldstone boson and $b=1,2$ corresponding to the physical charged Higgs bosons.
Notice that the rotation matrix $R_{ab}$ not only diagonalizes the charged scalar sector, 
but also brings us to a Higgs basis: $v_a = R_{a0} v$.
Then, in terms of the physical fields, the interactions have the form
\begin{equation}
-{\cal L}_{ch.} = \bar u^{\rm phys.} \left( \tilde \Gamma_b \cdot P_R - \tilde \Delta^\dagger_b \cdot P_L\right) d^{\rm phys.} H_b^+ + h.c.\label{Yukawa-charged-2}
\end{equation}
Here, $P_L = (1-\gamma^5)/2$ and $P_R = (1+\gamma^5)/2$ are the chiral projectors and
\begin{equation}
\tilde \Gamma_b = V_{uL}^\dagger \cdot \Gamma_a R_{ab} \cdot V_{dR}\,, \quad 
\tilde\Delta^\dagger_b =  V_{uR}^\dagger \cdot \Delta^\dagger_a R_{ab} \cdot V_{dL} \,.
\end{equation}
Both matrices have the following generation structure:
\begin{equation}
\tilde \Gamma, \tilde\Delta^\dagger \sim \mmmatrix{ud}{us}{ub}{cd}{cs}{cb}{td}{ts}{tb}\,.
\end{equation}
One can also explicitly factor out the CKM matrix $V$ and represent these interaction matrices as 
\begin{equation}
\tilde \Gamma_b = V \cdot V_{dL}^\dagger \Gamma_a R_{ab} V_{dR}\,, \quad 
\tilde\Delta^\dagger_b =  V_{uR}^\dagger \Delta^\dagger_a R_{ab} V_{uL} \cdot V\,.
\end{equation}
Before we proceed with CP4 3HDM, it is instructive to see how Eq.~\eqref{Yukawa-charged-2} 
simplifies in models with natural flavor conservation (NFC) \cite{Albright:1979yc,Branco:1985pf,Grossman:1994jb}. 
In these cases, only one structure $\Gamma$ is responsible for down quark mass matrix
and only one structure $\Delta$ gives rise to the up-quark mass matrix.
Therefore, the charged Higgs interactions become
\begin{equation}
-{\cal L}_{ch.}^{(NFC)} =  {\sqrt{2} \over v} \bar u^{\rm phys.} \left( Y_b V D_d \cdot P_R + X_b D_u V \cdot P_L\right) d^{\rm phys.} H_b^+ + h.c.,\label{Yukawa-charged-NFC-1}
\end{equation}
where $D_d = \mathrm{diag}(m_d, m_s, m_b)$ and $D_u = \mathrm{diag}(m_u, m_c, m_t)$.
Here, $X_b$ and $Y_b$, $b=1,2$ are {\em numbers}, not matrices, 
and they depend on the particular type of the NFC realization. 
If one needs to extract a specific flavour pair, then the coupling
becomes proportional to the corresponding CKM matrix element.
In particular, the strongest coupling is $\bar{t}b H^+$ coming from
\begin{equation}
-{\cal L}_{tbH^+}^{(NFC)} =  {\sqrt{2} \over v} V_{tb}\, \bar t^{\rm phys.} \left( Y_b m_b P_R + X_b m_t P_L\right) b^{\rm phys.} H_b^+ + h.c.\label{Yukawa-charged-NFC-2}
\end{equation}
The coefficients $X$ and $Y$ depend on the models and can also be constrained 
from the experiment, see for example \cite{Grossman:1994jb}.
Within 2HDM with NFC, there is only one charged Higgs, and its coefficient
can be related with the angle $\beta$. For example, Type I 2HDM leads to $X = -Y= \cot\beta$, 
while in Type II 2HDM $X = \cot\beta$, $Y = \tan\beta$.

CP4 3HDM does {\em not} possess the NFC property.
Individual Yukawa structures $\Gamma_a$ and $\Delta_a$ cannot produce viable
quark mass matrices. It is crucial that several structures sum up to produce the mass matrices $M_d$ and $M_u$.
Therefore, in CP4 3HDM, we do not expect the charged Higgs coupling matrices $\tilde \Gamma_b$
and $\tilde \Delta^\dagger_b$ to always bear the CKM structure.
In fact, as we will see below, there exist parameter space points with non-2HDM-like patterns.
Since existing experimental searches are partially motivated by Type I or Type II 2HDM predictions,
these points may avoid existing constraints. 
At the same time, they will show up strong in novel, non-canonical final states,
and can be checked in future.

\subsection{Decays $t\to H^+ d_j$ and $H^+ \to u_i\bar d_j$}

If the charged Higgs boson $H^+$ is sufficiently light, the top-quark can decay as $t \to d_j H^+$.
This new channel leads to two effects: a modification of the total top-quark width with respect to the SM value,
and the appearance of a novel final-state signal, which depends on the $H^+$ decay preferences.

At tree level, the decay width $t \to d_j H^+$, where $d_j = (d, s, b)$ with masses $m_j$, can be written as
\begin{eqnarray}
\Gamma_{t \to d_j H^+} &=& {\sqrt{\lambda(m_t^2,m_{H^+}^2,m_j^2)} \over 32 \pi m_t^3} 
\left[(m_t^2+m_j^2-m_{H^+}^2) (|\tilde \Gamma_{tj}|^2 + |\tilde \Delta^\dagger_{tj}|^2) 
- 4 m_t m_j \Re (\Gamma_{tj}\Delta_{jt})\right]\,,\nonumber\\
&\approx & {|\tilde \Gamma_{tj}|^2 + |\tilde \Delta^\dagger_{tj}|^2 \over 32 \pi} m_t 
\left(1 - {m_{H^+}^2 \over m_t^2}\right)^2\,.\label{tHb-1}
\end{eqnarray}
Here, we introduced the function 
\begin{equation}
\lambda(m_t^2,m_{H^+}^2,m_j^2) = m_t^4 + m_{H^+}^4 + m_j^4 - 2 m_{H^+}^2m_t^2 - 2m_{H^+}^2m_j^2 - 2m_t^2m_j^2\,.
\end{equation}
The second line \eqref{tHb-1} corresponds to neglecting the light quark mass $m_j \to 0$.
Notice that $\Gamma_{tj}$ and $\Delta_{jt}$ denote here individual matrix entries, not the entire matrices.

To get a qualitative estimate of the importance of this channel, 
let us compare this contribution with the SM top decay width,
which at tree level and in the approximation $m_b = 0$ has the form
\begin{equation}
\Gamma_{SM} = {G_F m_t^3 \over 8 \sqrt{2} \pi}\left(1 - {m_W^2 \over m_t^2}\right)^2
\left(1 + 2{m_W^2 \over m_t^2}\right)\,.\label{tWb-SM}
\end{equation}
Since $2\sqrt{2} G m_t^2 = 2 m_t^2/v^2 \approx 1$, 
one gets $\Gamma_{SM}/m_t \approx 1/(32\pi)$ times the brackets of Eq.~\eqref{tWb-SM},
which yields $\Gamma \sim 1$ GeV.
Comparing the two decay widths, we see that the competition is essentially 
between $|\tilde \Gamma_{tj}|^2 + |\tilde \Delta^\dagger_{tj}|^2$
and $(1+2m_W^2/m_t^2)/4 \approx 0.4$,
corrected by the corresponding phase space factors.
So, if $\sum_j (|\tilde \Gamma_{tj}|^2 + |\tilde \Delta_{tj}^\dagger|^2)$ stays well below $0.1$,
the contribution of the new channels to the top-quark width can be neglected. 
If $\sum_j (|\tilde \Gamma_{tj}|^2 + |\tilde \Delta_{tj}^\dagger|^2) \sim 1$,
we must check whether the phase space factor in Eq.~\eqref{tHb-1} 
provides sufficient suppression of the new contribution to the decay width.

Once $H^+_b$ are produced in top decays, they subsequently decay into
pairs of lighter quarks $H^+ \to u_i \bar{d}_j$, where $u_i = (u, c)$
and $d_j = (d, s, b)$. 
	Keeping only one of the quark masses $m$ non-zero
(which is done only to track the threshold shift in the $H^+ \to u_i \bar{b}$ decays), 
we calculate the corresponding decay width as
\begin{equation}
\Gamma_{ij} = \Gamma(H^+ \to u_i\bar d_j) = {N_c m_{H^+} \over 16 \pi} \left(1 - {m^2 \over m_{H^+}^2}\right)^2 
(|\tilde \Gamma_{ij}|^2 + |\tilde \Delta^\dagger_{ij}|^2)\,,
\label{Gamma-ij}
\end{equation}
where $N_c = 3$ is the number of quark colors.
Since we focus on cases with sufficiently light charged Higgses, $m_{H^+}\sim 90-160$ GeV,
which were dominant in the numerical scan of \cite{Ferreira:2017tvy},
there are no $H^+ \to W^+ H$ decays, with $H$ being either SM 
or extra neutral Higgses.

We would like to stress an important difference between the $H^+ \to q\bar q$ decays, for which no SM counterpart exists,
and the charged Higgs channel of the $t$ decays.
Since the additional Higgses are assumed to be leptophobic, 
$H^+$ decays only to quark pairs. 
The branching ratios of individual channels arise not from the absolute magnitudes 
of the entries $\tilde \Gamma_{ij}$ and $\tilde\Delta_{ji}$ but from 
the competition among all the entries.
Therefore, if an individual entry $\tilde \Gamma_{ij} \ll 1$,
it nevertheless can be the dominant decay channel $H^+$ if all the other entries are even smaller.

\section{Numerical results}\label{section-numerical}

\subsection{Top quark width}

For numerical calculations, we use the parameter space points, including the matrices 
$\tilde \Gamma_{ij} $ and $\tilde \Delta^\dagger_{ij}$ produced by the 
numerical scan of Ref.~\cite{Ferreira:2017tvy}.
The total statistics of points passing all criteria used in \cite{Ferreira:2017tvy}
is: 9 points for case $(B_1, B_1)$, 10 points for case $(B_2, B_2)$,
48 points for case $(B_1, B_3)$.
	Among these 67 point, 5 did not contain charged Higgses lighter than the top quark.
These five parameter space are not constrained by the present analysis.
From the remaining 62 points, one third (21 points) contained two charged Higgses ligher than 170 GeV.
When analyzing them, we will plot branching ratios for each of the two Higgses.
The remaining points contained only one charged Higgs lighter than the top quark.

We start by checking the charged Higgs contributions to the total top quark decay width.
We calculate $\Gamma_t = \Gamma_{SM} + \sum_j\Gamma(t \to H_{1,2}^+d_j)$,
where the NLO (EW) + NNLO (QCD) Standard Model result $\Gamma_{SM} = 1.322$ GeV is taken from \cite{Gao:2012ja}, 
while $\sum_j\Gamma(t \to H_{1,2}^+d_j)$
is computed from Eq.~\eqref{tHb-1} summed over all the relevant channels. 
The latest PDG total top-quark decay width is $\Gamma_t = 1.42 {}^{+0.19}_{-0.15}$ GeV \cite{Zyla:2020zbs}.
It is consistent with the SM calculation but also leave some room for New Physics contributions
with branchng ratios at the level of tens of percents. 

\begin{figure}[!htb]
\begin{center}
	\includegraphics[width=0.8\textwidth]{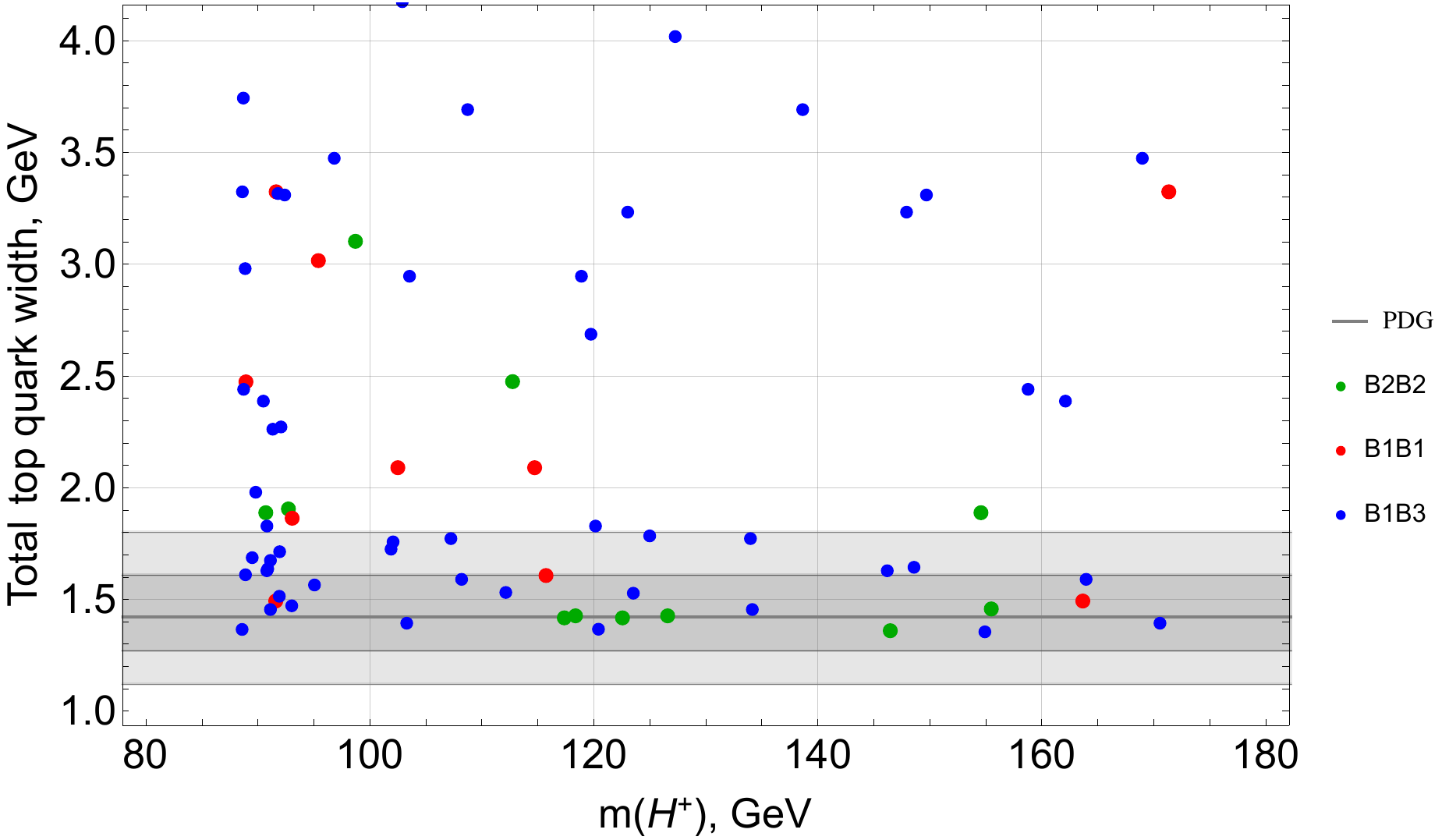}
\end{center}	
	\caption{The total top-quark width calculated as the SM contribution plus all the $t \to H^+_a d_j$ contributions 
		for all the parameter space points used. Different colors correspond to the three Yukawa sector combinations.
		The shaded regions cover $\pm 1\sigma$ and $\pm 2\sigma$ regions around the central PDG value.}
\label{Fig1}
\end{figure}

The results of this comparison are shown in Fig.~\ref{Fig1},
where different colors correspond to cases $(B_1, B_1)$, $(B_2, B_2)$, and $(B_1, B_3)$.
If a parameter space point contains two light charged Higgses, we plot two points corresponding to the two $M_{H^\pm}$ values,
with the common $\Gamma_t$ summed over both charged Higgs contributions.
We allow for at most a $2\sigma$ upward deviation from the central experimental value,
which implies that we accept a point if the partial decay width $t \to H^+_a d_j$, when summed over both kinematically allowed
charged Higgses $H_a^+$ and over all down-type quarks $d_j$, is less than 0.5 GeV.
As it can be seen on this plot, many points lead to a significant charged Higgs contribution to $\Gamma_t$
and are excluded by this check (in fact, there exist a few points with $\Gamma_t > 4$ GeV). 
Nevertheless, roughly half of all parameter space points survive this check.

\subsection{Branching ratios}

Now we turn to the searches for light charged Higgses via their production
in top-quark decays $t \to H^+ b$ and subsequent hadronic decays
$H^+ \to u_i \bar d_j$, where $u_i \bar d_j = c\bar{s}$ \cite{Aad:2013hla,Khachatryan:2015uua,Sirunyan:2020aln} and 
$c\bar{b}$ \cite{Sirunyan:2018dvm}.
The results of these searches are presented as upper limits on the top-quark branching fraction $Br(t\to b H^+) < p$ 
under the assumption that the corresponding $H^+$ decay is fully dominated by the selected hadronic channel:
$Br(H^+ \to c\bar s) = 100\%$ in \cite{Aad:2013hla,Khachatryan:2015uua,Sirunyan:2020aln} and
$Br(H^+ \to c\bar b) = 100\%$ in \cite{Sirunyan:2018dvm}.
In our case, $H_{1,2}^+$ have several decay channels. 
Therefore, we present these results as upper limits on the {\em product} of branching ratios
corresponding to production and decay of $H^\pm$ of the specific channel used in the experiment:
\begin{equation}
Br(t\to b H^+) \times Br(H^+ \to u_i \bar d_j)  < p\,.\label{BrBr}
\end{equation}
The value of $p$ depends on the channel and on the charged Higgs mass. 
The strongest limits
correspond to $p = 0.25\%$ in the $c\bar s$ channel \cite{Sirunyan:2020aln}
and $p = 0.5\%$ in the $c\bar b$ channel \cite{Sirunyan:2018dvm}, both established by the CMS experiment.
Although for the $H^+$ masses close to $m_W$ or to $m_t$ the limits are weaker,
we used the most conservative (the strongest) limits for all the charged Higgs masses.

\begin{figure}[!htb]
\centering
\includegraphics[width=0.7\textwidth]{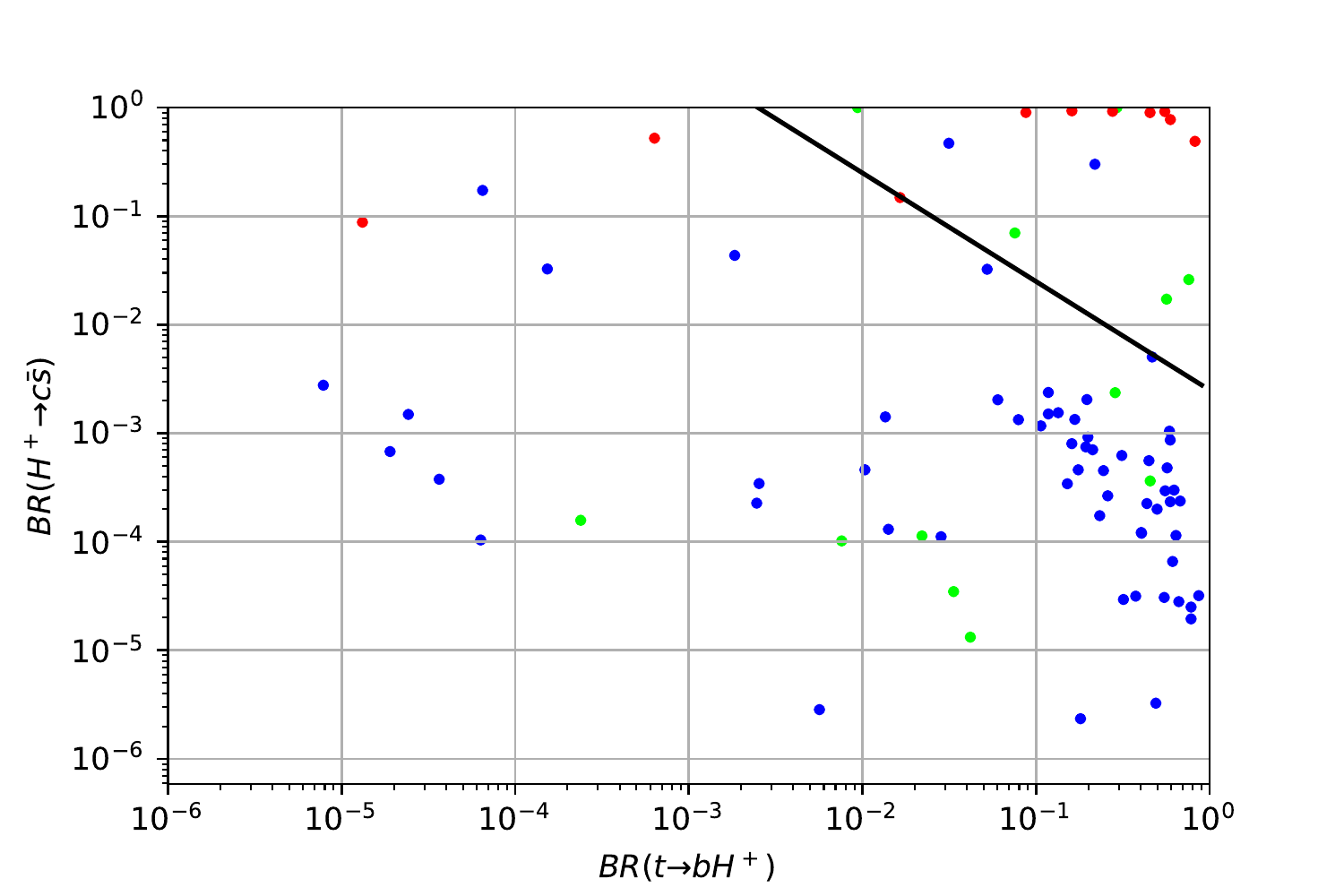}\\
\includegraphics[width=0.7\textwidth]{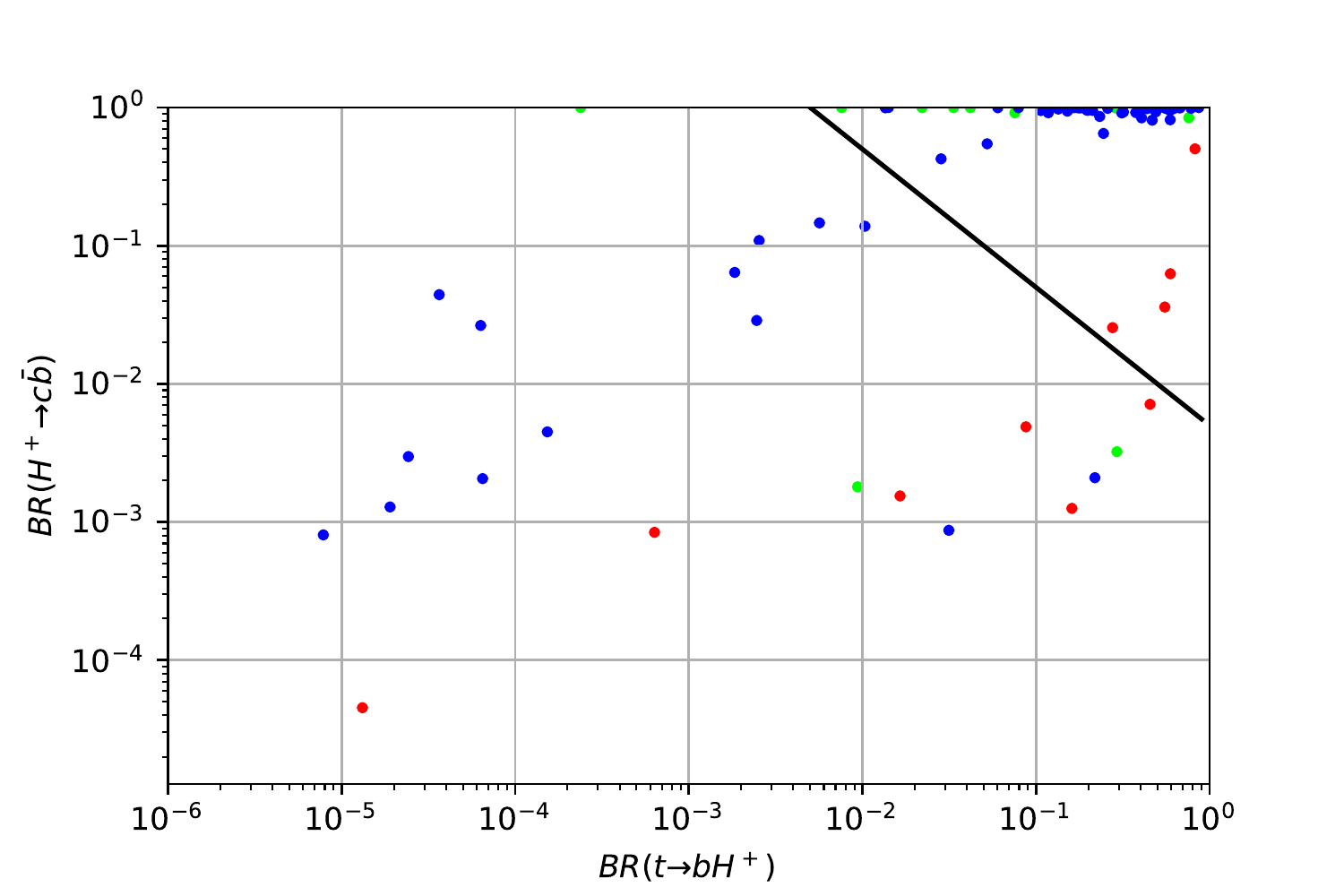}
\caption{Branching ratios $Br(H^+ \to c\bar s)$ (top) and $Br(H^+\to c\bar b)$ (bottom)
vs. $Br(t \to H^+ b)$. The colors encoding the Yukawa combinations
$(B_1, B_1)$, $(B_2, B_2)$, and $(B_1, B_3)$ are the same as in Fig.~\ref{Fig1}.
The oblique lines corresponds to the LHC upper limits \eqref{BrBr}
with $p = 0.25\%$ for $c\bar s$ \cite{Sirunyan:2020aln} and 
$p = 0.5\%$ for $c\bar b$ \cite{Sirunyan:2018dvm}.}\label{Fig2}
\end{figure}

In Fig.~\ref{Fig2} we compare the prediction for the parameter space points 
found in Ref.~\cite{Ferreira:2017tvy} with the experimental constraints \eqref{BrBr}.
We plot the points on the plane 
$Br(H^+ \to c\bar s)$ (the upper plot) or $Br(H^+ \to c\bar b)$ (the lower plot) 
vs $Br(t\to b H^+)$ and draw the line corresponding to the upper limit
\eqref{BrBr}. Only points lying below the line on {\em both} plots can be considered as passing the check.
If a parameter space point contains two light charged Higgses, 
we analyze their signals individually and plot them as separate points on these plots.
In this case, in order for a model to pass the check, both charged Higgses must stay below
the lines on both plots.

As can be immediately seen, the vast majority of the points found in \cite{Ferreira:2017tvy}
fail this check for $H^+\to c\bar b$ (case $(B_1, B_3)$, blue points) 
or for $H^+\to c\bar s$ (cases $(B_1, B_1)$ and $(B_2, B_2)$, red and green points). 
Also, in some cases, when a point appears on both plots below the line, it corresponds to the {\em second} charged Higgs
of the model lying close to the $t \to H_2^+b$ decay threshold. 
However, the first charged Higgs of the same model usually leads to large branching ratios
and, therefore, the model is ruled out. 

\subsection{Notable parameter space points}

Among the 62 candidate models reported in \cite{Ferreira:2017tvy} with at least one charged Higgses lighter than the top quark, 
only one model passed all our checks.
This is a $(B_1, B_3)$ model with a very remarkable quark interaction patterns. 
This point contains only one light charged Higgs of mass $m(H_1^+) = 155$ GeV, which is only slightly ligher than the top quark. 
At the same time, the largest top coupling is only $(\tilde \Delta^\dagger)_{tb} = 0.34$,
so that $Br(t \to H^+b)$ is well below 1\%. 
As for the subsequent decay of $H_1^+$, its couplings with quarks exhibit exotic patterns across generations:
\begin{equation}
\tilde \Gamma \approx \mmmatrix{0.60}{0.06}{0.21}{0.14}{0.08}{0.40}{0.007}{0.005}{8.5}\times 10^{-3}\,,\quad
\tilde\Delta^\dagger \approx \mmmatrix{0.0008}{0.0011}{0.168}{0.006}{0.0003}{0.070}{0.039}{0.003}{0.34}\,.\label{point60}
\end{equation}
The numbers indicated refer to the absolute values of matrix entries.
By inspecting the first two rows in these matrices, we conclude that the dominant
decay mode is neither $H^+ \to c\bar s$ nor $H^+\to c\bar b$ but $H^+ \to u\bar b$ with the branching ratio of about 90\%. 
No experimental search exists for such final state of the charged Higgs decay. Even if the CMS results \cite{Sirunyan:2018dvm} 
can be recast in an equally strong constraint on $H^+ \to u\bar b$ as on $H^+ \to c\bar b$,
this point would still pass the test thanks to the small $Br(t \to H^+b)$.

We also found other points which were close to satisfying all our checks.
Perhaps, the most intriguing example is given by a $(B_1, B_1)$ model which contains 
two light charged Higgses with masses $m(H_1^+) = 103$ GeV
and $m(H_2^+) = 115$ GeV.
Their strongest couplings to quarks come from the following matrices $\tilde\Delta^\dagger$:
\begin{equation}
\tilde\Delta^\dagger_1 \approx \mmmatrix{2\cdot 10^{-5}}{0.187}{0.003}{4\cdot 10^{-5}}{0.194}{0.008}{2\cdot 10^{-4}}{0.989}{0.042}\,,\quad
\tilde\Delta^\dagger_2 \approx \mmmatrix{5\cdot 10^{-5}}{9\cdot 10^{-4}}{1.4\cdot 10^{-5}}{0.0065}{0.0027}{1\cdot 10^{-4}}{0.035}{0.015}{0.252}\,.
\end{equation}
The two charged Higgses display very different preferences. The first one, $H_1^+$, is produced in the $t \to H^+ s$ decay 
and because of that it easily avoids experimental constraints. 
The second one, $H_2^+$, has mildly suppressed branching ratios and barely passes the decay constraints; 
it is visible in the top plot of Fig.~\ref{Fig2} as the red dot right on the line. 
However, due to the large $H_1^+ts$ coupling, the total top quark decay width is about 2.1 GeV,
which conflicts the measurements.

These observations highlight the necessity of simultaneously checking all three observables: $\Gamma_t$,
$Br(t\to bH^+ (\to c\bar b))$ and $Br(t\to bH^+ (\to c\bar s))$. They also confirm the initial expectations
that models with very exotic, non-2HDM-like patterns of $H^+q\bar q$ couplings can arise in CP4 3HDM.


\section{Discussion and conclusions}

In this work, we continued exploration of a unique three-Higgs-doublet model
equipped with a higher-order $CP$ symmetry, which was suggested first in \cite{Ivanov:2015mwl}.
We used the results of the parameter space scan performed in \cite{Ferreira:2017tvy} and focused on the charged Higgs bosons.
It turns out that almost all parameter space points which passed the electroweak precision and flavor constraints
of \cite{Ferreira:2017tvy} contain one or two charged Higgses lighter than the top quark.
As a result, a new top decay channels open up, such as $t \to H^+ b$ with subsequent hadronic decay of $H^+$.

Unlike other studies of the light charged Higgses in 3HDMs such as 
\cite{Akeroyd:2012yg,Akeroyd:2016ssd,Akeroyd:2018axd,Akeroyd:2019mvt,Chakraborti:2021bpy},
we do not---and cannot---assume the natural flavor conservation within CP4 3HDM.
Thus, we can suspect that many points emerging from the scan of \cite{Ferreira:2017tvy} could be in conflict with experimental data.

After Ref.~\cite{Ferreira:2017tvy} was published, new LHC results on light charged scalars searches 
appeared. In particular, the two CMS searches \cite{Sirunyan:2018dvm,Sirunyan:2020aln} of light charged Higgses
emerging from top decays $t \to H^+ b$ and decaying hadronically to $Br(H^+ \to c\bar s)$ or $Br(H^+ \to c\bar b)$
placed subpercent level upper limits on the relevant branching ratios.
In this work we took these new data into account and checked whether the parameter space points 
considered viable in \cite{Ferreira:2017tvy} were compatible with these new results.
We also took into account the updated value of the top quark total width,
which places an upper bound on any non-standard decay of the top.

Out of 67 parameter space points borrowed from \cite{Ferreira:2017tvy}, five contained no charged Higgses
lighter than top. These points remain viable and must be subjected to other experimental constraints.
Among the remaining 62 points, only one passed all the three experimental constraints ($\Gamma_t$, $c\bar{s}$ and $c\bar{b}$ decays).
This point avoided the experimental constraints because of its non-2HDM-like pattern
of the $H^+$-quark couplings, see the matrices \eqref{point60}: the dominant decay channel is $H^+ \to u\bar b$
which was not searched for in experiment. 

We also observed other examples where peculiar $H^+$-quark patterns allowed the charged Higgses to avoid two tests
and only moderately fail the third one. Thus, such exotic patterns defying the 2HDM-based intuition are not exceptional
and represent an intriguing feature of the CP4 3HDM.

It is interesting to check whether additional parameters space scans of the CP4 3HDM can identify other 
benchmark models with unusual charged Higgs patterns.
To this end, we want to mention that the numerical scan of \cite{Ferreira:2017tvy} used two additional assumptions:
the 125 GeV was identified with the lightest neutral scalar, and the exact alignment was assumed in the scalar sector.
The results of the present work show that these assumptions within CP4 3HDM tend to conflict with the data.
By relaxing the alignment assumption, one can obtain many more viable points within the CP4 3HDM.
Although there are indications that the true decoupling regime cannot be achieved within the spontaneously broken CP4 3HDM
\cite{Carrolo:2021euy}, one can still hope to generate benchmark models with heavier additional scalars which
satisfy all present collider constraints.
There remains much to explore within the CP4 3HDM.

\vspace{0.5cm}

{\bf Acknowledgments} 
We are grateful to Hugo Serôdio for providing us with numerical values of the parameter 
space points presented in \cite{Ferreira:2017tvy}.
We also thank Andrew Akeroyd for useful comments.
This work was partially supported by the National Science Center, Poland, via the project Harmonia (UMO-2015/18/M/ST2/00518).

\appendix

\section{CP4 symmetric Yukawa sectors}

The quark Yukawa lagrangian \eqref{Yukawa-general} can be made CP4 symmetric,
if the CP4 transformation acts on the fermion fields in a non-trivial way:
\begin{equation}
\psi_i \toCP Y_{ij} \psi_j^{CP}\,, \quad\mbox{where} \quad \psi^{CP} = \gamma^0 C \bar\psi^T\,.
\label{fermion-GCP-again}
\end{equation}
For each sector, $q_L$, $u_R$, and $d_R$, one can use its own $Y_{ij}$.
Within each sector, there always exists a basis in which the corresponding matrix $Y$ 
takes the form
\begin{equation}
Y = \mmmatrix{0}{e^{i\alpha}}{0}{e^{-i\alpha}}{0}{0}{0}{0}{1}\,,\label{matrix-Y}
\end{equation}
with parameters $\alpha$ which can be different for the three sectors.
The simultaneous solution of the consistency equations leads to one four possible options
for these matrices labeled in \cite{Ferreira:2017tvy} cases $A$, $B_1$, $B_2$, $B_3$:
\begin{itemize}
	\item
	Case $A$: $\alpha_L = \alpha_d = 1$, giving
	\begin{equation}
	\Gamma_1 = \mmmatrix{g_{11}}{g_{12}}{g_{13}}%
	{g_{12}^*}{g_{11}^*}{g_{13}^*}%
	{g_{31}}{g_{31}^*}{g_{33}}\,,\quad
	\Gamma_{2,3} = 0\,.\label{caseA}
	\end{equation}
	\item
	Case $B_1$: $\alpha_L = \pi/2$, $\alpha_d = 0$, giving
	\begin{equation}
	\Gamma_1 = \mmmatrix{0}{0}{0}{0}{0}{0}{g_{31}}{g_{31}^*}{g_{33}}\,,\quad
	\Gamma_2 = \mmmatrix{g_{11}}{g_{12}}{g_{13}}{g_{21}}{g_{22}}{g_{23}}{0}{0}{0}\,,\quad
	\Gamma_3 =  \mmmatrix{-g_{22}^*}{-g_{21}^*}{-g_{23}^*}{g_{12}^*}{g_{11}^*}{g_{13}^*}{0}{0}{0}\,.
	\label{caseB1}
	\end{equation}
	\item
	Case $B_2$: $\alpha_L = 0$, $\alpha_d = \pi/2$, giving
	\begin{equation}
	\Gamma_1 = \mmmatrix{0}{0}{g_{13}}{0}{0}{g_{13}^*}{0}{0}{g_{33}}\,,\quad
	\Gamma_2 = \mmmatrix{g_{11}}{g_{12}}{0}{g_{21}}{g_{22}}{0}{g_{31}}{g_{32}}{0}\,,\quad
	\Gamma_3 =  \mmmatrix{g_{22}^*}{-g_{21}^*}{0}{g_{12}^*}{-g_{11}^*}{0}{g_{32}^*}{-g_{31}^*}{0}\,.
	\label{caseB2}
	\end{equation}
	\item
	Case $B_3$: $\alpha_L = \pi/2$, $\alpha_d = \pi/2$, giving
	\begin{equation}
	\Gamma_1 = \mmmatrix{g_{11}}{g_{12}}{0}{-g_{12}^* }{g_{11}^*}{0}{0}{0}{g_{33}}\,,\quad
	\Gamma_2 = \mmmatrix{0}{0}{g_{13}}{0}{0}{g_{23}}{g_{31}}{g_{32}}{0}\,,\quad
	\Gamma_3 = \mmmatrix{0}{0}{-g_{23}^*}{0}{0}{g_{13}^*}{g_{32}^*}{-g_{31}^*}{0}\,.
	\label{caseB3}
	\end{equation}
\end{itemize}
All parameters apart from $g_{33}$ can be complex in each cases. Notice also that in all cases
the matrices $\Gamma_{2,3}$ are expressed in terms of the same complex parameters and
have the same textures.

The same list of cases exists for the up-quark sector. When constructing a viable model, 
we can combine different cases for down and up quarks, making sure that the transformation
properties of the left-handed doublets (defined by $\alpha_L$) are the same.
Therefore, we get two series of possible CP4 3HDM Yukawa sectors:
\begin{eqnarray}
\alpha_L = 0: && (A, A),\ (A, B_2),\ (B_2,A),\ (B_2, B_2),\\
\alpha_L = \pi/2: && (B_1, B_1),\ (B_1, B_3),\ (B_3, B_1),\ (B_3,B_3).
\label{combininig-cases-again}
\end{eqnarray}

\end{document}